\documentclass[prb,twocolumn,superscriptaddress,showpacs,amsmath,amssymb]{revtex4}
\usepackage{amsfonts}
\usepackage{bbm}
\usepackage{graphicx}% Include figure files
\usepackage{dcolumn}% Align table columns on decimal point
\usepackage{bm}% bold math

\begin{document}
\title{Spin-flip reflection at the normal metal-spin superconductor interface}

\author{Peng Lv}
\affiliation{International Center for Quantum Materials, School of Physics, Peking University, Beijing 100871, China}

\author{Ai-Min Guo}
\affiliation{Department of Physics, Harbin Institute of Technology, Harbin 150001, China}

\author{Huaiyu Li}
\affiliation{International Center for Quantum Materials, School of Physics, Peking University, Beijing 100871, China}

\author{Chunxiao Liu}
\affiliation{International Center for Quantum Materials, School of Physics, Peking University, Beijing 100871, China}

\author{X. C. Xie}
\affiliation{International Center for Quantum Materials, School of Physics, Peking University, Beijing 100871, China}
\affiliation{Collaborative Innovation Center of Quantum Matter, Beijing 100871, China}

\author{Qing-Feng Sun}
\email[]{sunqf@pku.edu.cn}
\affiliation{International Center for Quantum Materials, School of Physics, Peking University, Beijing 100871, China}
\affiliation{Collaborative Innovation Center of Quantum Matter, Beijing 100871, China}

\begin{abstract}
We study spin transport through a normal metal-spin superconductor junction.
A spin-flip reflection is demonstrated at the interface,
where a spin-up electron incident from the normal metal can be reflected as a
spin-down electron and the spin $2\times \hbar/2$ will be injected into the spin superconductor.
When the (spin) voltage is smaller than the gap of the spin superconductor, the spin-flip reflection determines
the transport properties of the junction.
We consider both graphene-based (linear-dispersion-relation) and quadratic-dispersion-relation normal metal-spin superconductor junctions in detail.
For the two-dimensional graphene-based junction,
the spin-flip reflected electron can be along the specular direction (retro-direction) when the incident and reflected electron locates in the same band (different bands).
A perfect spin-flip reflection can occur when the incident electron is normal to the interface, and the reflection coefficient is slightly suppressed for the oblique incident case.
As a comparison, for the one-dimensional quadratic-dispersion-relation junction,
the spin-flip reflection coefficient can reach 1 at certain incident energies.
In addition, both the charge current and the spin current under a charge (spin) voltage are studied.
The spin conductance is proportional to the spin-flip reflection coefficient when the spin voltage is less than the gap of the spin superconductor.
These results will help us get a better understanding of spin transport through the normal metal-spin superconductor junction.
\end{abstract}
\maketitle

\section{Introduction}

When a metal is coupled to a superconductor, the Andreev reflection occurs at the interface
between the metal and the superconductor,\cite{Andr}
where an electron incident from the normal metal is reflected as a hole at the interface and
a Cooper pair is generated in the superconductor.
The Andreev reflection determines the conductance of the normal metal-superconductor junction
when the voltage is smaller than the
superconducting gap since the normal tunneling cannot occur.
The quasi-particle current is transformed to a supercurrent at the metal-superconductor interface
and this process can be described by the scattering-matrix approach.\cite{sca1,sca2,sca3}
In usual metal-superconductor junctions, the Andreev reflected hole retraces
the path of the incident electron and thus this Andreev reflection is also called as Andreev retroreflection.
Besides, another kind of reflection, specular Andreev reflection, has also been reported in graphene-superconductor junctions.\cite{spe1,spe2,addr1}
Graphene is a single layer of carbon atoms arranged in a honeycomb lattice and has drawn great attention since its experimental realization.\cite{gra1,gra2,gra3}
Graphene has a unique band structure with a linear dispersion relation of low-lying excitations,
which gives rise to many peculiar properties.\cite{gra4}
Specular Andreev reflection occurs when the incident electron and the reflected hole locate, respectively, at the conduction and valence bands of graphene,\cite{spe1} whereas Andreev retroreflection occurs when the incident electron and the reflected hole locate at the same band.
Since then, many papers have studied graphene-based superconductor hybrid systems.\cite{addr1,gs1,gs2,gs3,gs4,gs5,addr2,addr3}

The conventional charge superconductivity can be regarded as a superfluid of electric charge,
where electrons form Cooper pairs in the superconductor and condense into the BCS ground
state.\cite{s1,book1,book2,book3}
Each Cooper pair carries an electric charge $2e$ and is spin singlet for s-wave pairing.
The charge superconductor can support dissipationless charge current at equilibrium, with resistance being zero.
Recently, a new quantum state, the spin superconductor, was proposed.\cite{ssc1,ssc2,ssc3,work1,lup1}
The spin superconductivity is a novel quantum state and can be viewed as a counterpart of
the charge superconductivity.
The charge superconductor is a superfluid of electric charge,
where the condensates are Cooper pairs as mentioned above.
However, the spin superconductor is a superfluid of spin, and can be formed by condensed bosons in sufficiently low temperatures.
The bosons are electrically neutral with their spins being nonzero.
The spin superconductor can carry dissipationless flow of spin current, with spin resistance being zero.
On the other hand, the spin superconductor is a charge insulator and the charge current cannot flow through it. The spin superconductivity may exist in spin-polarized triplet exciton systems of graphene,\cite{ssc1,ssc2,ssc3} $^3$He-$B$ superfluidity,\cite{He1,He2} Bose-Einstein condensate of magnetic atoms,\cite{NP1-111,nature471-83,PRL114-070401} Bose-Einstein condensate of magnons and spinons,\cite{add1,add2,add3} and so on.
The excitons are charge neutral with their spin being either singlet or triplet, and the condensation of excitons have been
realized in an electron-hole bilayer system.\cite{PRL68-1196,PRL73-304} A pure spin system described by quantum Heisenberg spin model can also support spin
superconductivity, as long as some effective bosonic degrees of freedom such as triplons exist.\cite{add1}
We also notice the close relation between the spin superconductivity and the spin superfluidity.\cite{ssc3,spin-gra}
Several methods have been proposed to realize dissipationless spin current in both ferro- and antiferro-magnetic insulators,\cite{ssc1,Spin1,Spin2,Spin3}
and the phenomenological two-fluid theory or the Ginzburg-Landau theory has been well established.\cite{work1,lup1,twofluid,work2}

As an analogy of the Andreev reflection at the metal-superconductor interface,
it is natural to study the reflection characteristics of a metal-spin superconductor interface.
The condensates of a charge superconductor are Cooper pairs, while they are electron-hole (e-h) pairs in a spin superconductor.
The electron-like spin-up carriers and the hole-like spin-up carriers attract each other and form e-h pairs due to the Coulomb interaction.
The Bose-Einstein condensate of these charge neutral spin-triplet e-h pairs form the ground state of a spin superconductor.
When an electron is injected from a normal metal to a spin superconductor,
it should not be reflected as a hole, since the condensates of e-h pairs are charge
neutral and don't need extra electric charge for their formation.
However, because the condensates carry nonzero spin angular momentum,
an injected electron should be spin-flip reflected for the formation of a new spin-triplet e-h pair in the spin superconductor.
Hence, at the metal-spin superconductor interface, the reflection occurs on the spin degree of freedom instead of
the charge degree of freedom. This could give rise to a spin current when a spin voltage is applied between the metal and the spin superconductor.

Motivated by this analogy, we carry out a theoretical study of transport properties of a normal metal-spin superconductor junction.
By using the scattering-matrix approach, the spin-flip reflection coefficient and the differential spin conductance are obtained.
Both linear and quadratic dispersion cases are considered.
For the linear dispersion case (graphene-based normal-spin superconductor junction),
a normal incident electron will be perfectly spin-flip reflected at the interface, although the Fermi wave lengths mismatch at the two sides.
The spin-flip reflection will be slightly suppressed for the oblique incident case, since the Fermi wave lengths mismatch begins to take effect.
We also find that there exists either the spin-flip specular reflection or the spin-flip retro-reflection in the
graphene-based junction, which depends on whether the incident and reflected electrons locate in the same band or the different bands.
While for the quadratic-dispersion-relation normal metal-spin superconductor junction, the spin-flip reflection coefficient will be strongly affected by the potential barrier.
In addition, the spin transport through the junction is investigated.
The spin conductance is proportional to the spin-flip reflection coefficient when the spin voltage is less than the gap of the spin superconductor.

The rest of this paper is organized as follows.
In Sec. \ref{sec:Hamiltonian}, we show the model Hamiltonian of the normal metal-spin superconductor junction, considering both the linear and quadratic dispersion cases.
In Sec. \ref{sec:linear}, we investigate the spin-flip reflection at the graphene-based normal-spin superconductor interface, which possesses the linear dispersion relation.
As a comparison, in Sec. \ref{sec:quadratic}, the reflection is studied at the quadratic normal metal-spin superconductor interface. Finally, a brief summary is presented in Sec \ref{sec:conclusions}.

\section{\label{sec:Hamiltonian}model Hamiltonian}

In this section, we present the model Hamiltonian of the normal metal-spin superconductor junction
in real space, for both the linear and quadratic dispersion cases.
In a spin superconductor, the condensates are e-h pairs.\cite{ssc1,ssc3}
The e-h pairs are formed between electron-like spin-up carriers and hole-like spin-up ones by the Coulomb interaction. These e-h pairs condense into the ground state of a spin superconductor.
In the following, we elaborate it in detail.
First, we consider the linear dispersion case where the spin superconductor can be found in a ferromagnetic graphene.
Regarding a two-dimensional (2D) sheet of graphene in the $x-y$ plane,
the 2D massless Dirac Hamiltonian is given by\cite{gra4}
\begin{small}
\begin{align}\label{eq:1a}
H_{0}=& v_{F}\left( \begin{array}{cc}
0 &  p_{x}-ip_{y} \\
p_{x}+ip_{y} & 0 \\
\end{array} \right) \nonumber \\
=&\hbar v_{F}\left( \begin{array}{cc}
0 &  k_{x}-ik_{y} \\
k_{x}+ik_{y} & 0 \\
\end{array} \right)
=\hbar v_{F}k\left( \begin{array}{cc}
0 &  e^{-i\theta} \\
e^{i\theta} & 0 \\
\end{array} \right),
\end{align}
\end{small}
where $v_{F}\approx10^{6}\ \rm m/s$ is the Fermi velocity in graphene, which acts as the effective velocity of light.
$\bm{p}\equiv(p_x, p_y)$ is the momentum operator
in the $x-y$ plane.
$\theta$ is the angle between the wave vector $\bm{k}\equiv(k_x, k_y)$
and the $x$-axis such that $\tan{\theta}=k_{y}/k_{x}$. The Hamiltonian can be diagonalized by performing a unitary transformation
\begin{align}\label{eq:1b}
U=\frac{\sqrt{2}}{2}\left( \begin{array}{cc}
1 & 1 \\
e^{i\theta} & -e^{i\theta} \\
\end{array} \right).
\end{align}
Then,
\begin{align}\label{eq:1c}
H_{0}'=U^{\dagger}H_{0}U=\left( \begin{array}{cc}
\hbar kv_{F} & 0 \\
0 & -\hbar kv_{F} \\
\end{array} \right).
\end{align}
If graphene grows on a ferromagnetic
material\cite{F1,F2,F3} or is under an external magnetic field,\cite{M1} the spin degeneracy will be broken.
In order to describe a ferromagnetic graphene, the Hamiltonian should be expanded to a $4\times4$ matrix,
including both the spin and pseudospin degrees of freedom\cite{ssc1,ssc3}
\begin{align}\label{eq:1d}
H'=\left( \begin{array}{cc}
H_{0}'+M\tau_{0} & 0 \\
0 & H_{0}'-M\tau_{0}  \\
\end{array} \right),
\end{align}
where $M$ is the ferromagnetic exchange split energy and $\tau_{0}$ is the $2\times2$ identity matrix in pseudospin space. In the derivation presented below, we set $M<0$. Note that we have ignored the valley degree of freedom, because the two valleys are degenerated and the intervalley coupling is usually very small due to the well separation of the two valleys in $k$ space. The four corresponding energy bands are $\epsilon_{\tau\sigma}=\tau\hbar k v_{F}+\sigma M$, where $\tau=\pm$ denotes the pseudospin and $\sigma=(\uparrow,\downarrow)$ represents the real spin. We focus on the energy bands which are close to the Fermi energy $E_{F}=0$, i.e., $\epsilon_{+\uparrow}=\hbar kv_{F}+M$ and $\epsilon_{-\downarrow}=-\hbar kv_{F}-M$, because only the energy bands near the Fermi surface are relevant. For convenience, these two energy bands are called low-energy bands.
The mean-field approximation of the e-h Coulomb attraction interaction can induce
a pairing potential $\Delta$ between the two bands $\epsilon_{+\uparrow}$ and
$\epsilon_{-\downarrow}$.\cite{ssc1}
Then, the mean-field Hamiltonian (including the e-h attraction interaction) can be written as\cite{ssc1}
\begin{small}
\begin{align}\label{eq:1e}
H''=\left( \begin{array}{cccc}
\hbar kv_{F}+M & 0 & 0 & \Delta \\
0 & -\hbar kv_{F}+M & \Delta' & 0 \\
0 & \Delta' & \hbar kv_{F}-M & 0 \\
\Delta & 0 & 0 & -\hbar kv_{F}-M \\
\end{array} \right).
\end{align}
\end{small}
Here, we have added a pairing potential $\Delta'$ between the energy bands $\epsilon_{-\uparrow}$ and $\epsilon_{+\downarrow}$.
Usually $\Delta'\ll\Delta$ and we take $\Delta'=\Delta$ below for convenience of calculations.
Because the term involving $\Delta'$ is between the two high-energy bands $\epsilon_{-\uparrow}$ and $\epsilon_{+\downarrow}$, it has little effect on the transport properties around the Fermi surface.
Note that this Hamiltonian is consistent with Ref.[\onlinecite{ssc1}].

Next, the Hamiltonian is inverted back to its original form by taking a unitary transformation with $\tilde{U}=diag(U,U)$, i.e., $H_{SSC}=\tilde{U}H''\tilde{U}^{\dagger}$. The Hamiltonian of a spin superconductor is then given by
\begin{align}\label{eq:1f}
H_{SSC}=\left( \begin{array}{cc}
H_{0}+M\tau_{0} & \Delta\tau_{z} \\
\Delta\tau_{z} & H_{0}-M\tau_{0} \\
\end{array} \right),
\end{align}
where $\tau_{z}$ is the $z$-component of the Pauli matrices in pseudospin space. Note that this Hamiltonian is consistent
with the one when the graphene possesses the canted antiferromagnetic phase.\cite{H1,H2}
For the graphene-based normal-spin superconductor junction,
the graphene is in the spin superconducting phase when $x>0$
and is in the normal phase when $x<0$.
The Hamiltonian of the normal graphene at $x<0$ is
\begin{small}
\begin{align}\label{eq:1g}
H_{FMG}=\left( \begin{array}{cc}
H_{0}+(V+M_{L})\tau_{0} & 0 \\
0 & H_{0}+(V-M_{L})\tau_{0} \\
\end{array} \right).
\end{align}
\end{small}
Because the ferromagnetic graphene has a large magnetic momentum $M$ at $x>0$,
here we consider a small magnetic momentum $M_L$ $x<0$ also.
$V$ is a constant potential, which can be modulated by applying an external electric field or a gate voltage.
To summarize, the Hamiltonian of the graphene-based normal-spin superconductor junction can be expressed in a unified form as
\begin{small}
\begin{align}\label{eq:1h}
H=\left( \begin{array}{cc}
H_{0}+(M(x)+V(x))\tau_{0} & \Delta(x)\tau_{z} \\
\Delta(x)\tau_{z} & H_{0}+(-M(x)+V(x))\tau_{0} \\
\end{array} \right),
\end{align}
\end{small}
where $\Delta(x)=\Delta\Theta(x)$, $M(x)=M_L(1-\Theta(x))+M\Theta(x)$,
$V(x)=V(1-\Theta(x))$, and $\Theta(x)$ is the Heaviside step function.
We take $|M|\gg|M_{L}|$ in the calculation below.

For the quadratic dispersion case, we focus on a two-bands model and replace the linear low-energy bands $\epsilon_{+\uparrow}=\hbar kv_{F}+M$ and $\epsilon_{-\downarrow}=-\hbar kv_{F}-M$ in Eq.(\ref{eq:1e})
with a quadratic dispersion relation. The two-bands model Hamiltonian is then given by
\begin{align}\label{eq:1i}
H_{SSC}=\left( \begin{array}{cc}
\frac{p^2}{2m}+M & \Delta \\
\Delta & \frac{-p^2}{2m}-M \\
\end{array} \right).
\end{align}

At the normal side, we consider the quadratic metal, in which
the pairing potential $\Delta$ vanishes and the spin degeneracy remains.
The Hamiltonian is written as
\begin{align}\label{eq:1j}
H_{N}=\left( \begin{array}{cc}
\frac{p^2}{2m}+V & 0 \\
0 & \frac{p^2}{2m}+V \\
\end{array} \right).
\end{align}

Thus, the Hamiltonian of the quadratic normal metal-spin superconductor junction
can be written in a unified form as
\begin{small}
\begin{align}\label{eq:1k}
\hat{H}=\left( \begin{array}{cc}
\frac{p^2}{2m} & \Delta(x) \\
\Delta(x) & \frac{-p\cdot\mathrm{sgn}(x)p}{2m} \\
\end{array} \right)+V(x)+M(x)\sigma_{z}+V_0\delta(x).
\end{align}
\end{small}
Here, $V(x)=(1-\Theta(x))V$ is the potential energy at the normal side,
$M(x)=\Theta(x)M$ is the magnetic momentum of the spin superconductor,
and $V_0\delta(x)$ is the Schottky potential at the normal metal-spin superconductor interface.
Note that $\sigma_{z}$ is the $z$-component of the Pauli matrices in real spin space rather than pseudospin space.
We take $M<0$ and $|M|\gg\Delta$ in the calculation below, because the gap of the superconductor
is usually much smaller than the magnetic momentum.

\section{\label{sec:linear}Linear dispersion case}

\begin{figure}[h]
\includegraphics[width=0.9\columnwidth]{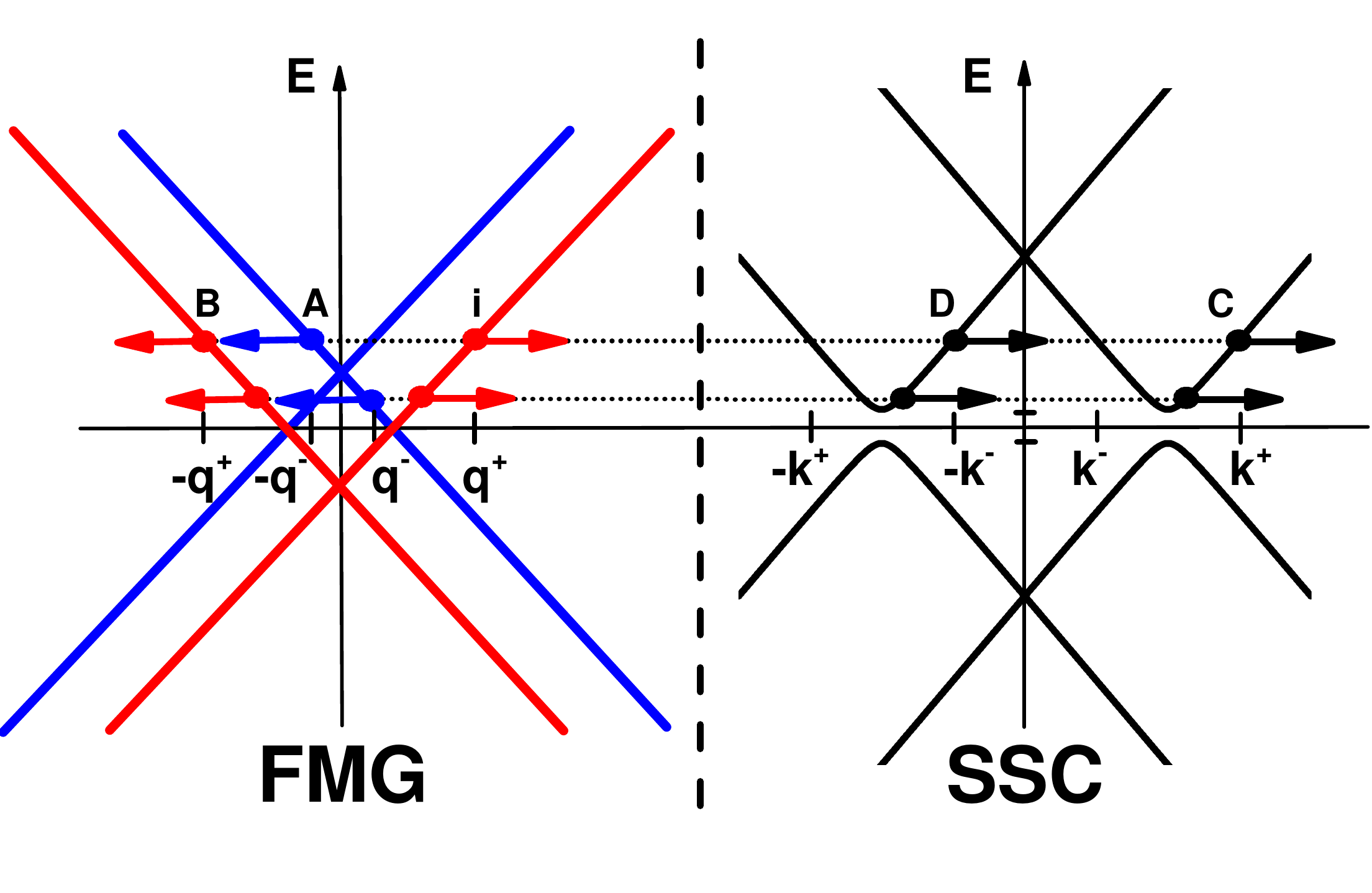}
\caption{(Color online) Band structures of a graphene with small magnetic momentum (FMG)
and of a spin superconductor (SSC).
In the graphene side, the red line denotes the dispersion spectrum of spin-up electrons and the blue line represents that of spin-down ones.
A spin-up electron i with wave vector $q^{+}$ is incident from the graphene side to the spin superconductor.
Spin-flip reflected electron A, normal reflected electron B, normal transmitted electron C, and
spin-flip transmitted electron D are represented by the filled circles, with the direction of motion indicated by the arrows.
There are two kinds of incident energies as shown in the figure. For one kind of incident energies, the incident electron
and the spin-flip reflected electron locate at the same band, whereas for the other kind of incident energies, they locate at
the conduction and valence bands.
} \label{fig1}
\end{figure}

In this section, we investigate the 2D scattering
at the graphene-based normal metal-spin superconductor interface.
The Hamiltonian corresponding to this case is shown in Eq.(\ref{eq:1h}).
We consider that the wave functions at both sides have the plane wave form, i.e.,
$\psi(x)=(a,b,c,d)^Te^{i\bm{q}\cdot\bm{r}}$.
For the normal graphene, we have the energy spectrum
\begin{subequations}
\begin{align}
\label{eq:2a1}
E_{\uparrow}(q_x,q_y)&=M_L+V\pm v_{F}\hbar\sqrt{q_x^2+q_y^2}, \\
\label{eq:2a2}
E_{\downarrow}(q_x,q_y)&=-M_L+V\pm v_{F}\hbar\sqrt{q_x^2+q_y^2},
\end{align}
\end{subequations}
as illustrated in Fig.\ref{fig1} (left side). $\bm{q}=(q_x, q_y)$ is the wave vector.
Because $M_{L}$ is set to be negative, the energy spectrum
of spin-up electrons will be shifted down.
For the spin superconductor, the energy spectrum can be obtained by diagonalizing the Hamiltonian and one gets
\begin{align}\label{eq:2b}
E=\pm\sqrt{\left(M\pm v_{F}\hbar\sqrt{q_x^2+q_y^2}\right)^2+\Delta^2},
\end{align}
where an energy gap $\Delta$ appears near the Fermi surface as shown in Fig.\ref{fig1} (right side).
In numerical calculations, we mainly study the quantum transport near the Fermi surface and thus the low-energy bands take effect
\begin{align}\label{eq:2c}
E=\pm\sqrt{\left(M+v_{F}\hbar\sqrt{q_x^2+q_y^2}\right)^2+\Delta^2}.
\end{align}

When a spin-up electron is injected from the left side to the spin superconductor, there exist two cases: (1) The incident electron and the spin-flip reflected electron locate in the same band;
(2) They locate, respectively, in the conduction and valence bands, as shown in Fig.\ref{fig1}.
In the following, we consider a 2D scattering problem and study the spin-flip reflection for both cases in detail.
There are three conservation laws in the present 2D scattering system.
First, there exists the conservation of energy due to the time translation invariance.
Then, the momentum projection $q_{y}$ should be invariant as a result of the translational invariance along the $y$-direction
(i.e., parallel to the interface).
Finally, the probability current is conserved along both the $x$- and $y$-directions.
Based on these conservation laws, the wave functions at both sides can be written.
When a spin-up electron i in the conduction band with energy $E>0$ is injected from the left side to the spin superconductor (see Fig.\ref{fig1}),
the spin-flip reflected electron A locates in the conduction/valence band of the spin-down energy spectrum. The wave function at the left side ($x<0$) is given by
\begin{small}
\begin{align}\label{eq:2d}
\psi(x,y)=&\left\{\frac{1}{\sqrt 2}\left(\begin{array}{c}
e^{-i\frac{\alpha^+}{2}} \\
e^{i\frac{\alpha^+}{2}} \\
0 \\
0 \\
\end{array}\right)e^{iq_{x}^{+}x}
+\frac{r_{\downarrow\uparrow}}{\sqrt 2}\left(\begin{array}{c}
0 \\
0 \\
e^{i\frac{\tau\alpha^-}{2}} \\
-e^{-i\frac{\tau\alpha^-}{2}} \\
\end{array}\right)e^{-i\tau q_{x}^-x}\right. \nonumber \\
&\left.+\frac{r_{\uparrow\uparrow}}{\sqrt 2}\left(\begin{array}{c}
e^{i\frac{\alpha^+}{2}} \\
-e^{-i\frac{\alpha^+}{2}} \\
0 \\
0 \\
\end{array}\right)e^{-iq_{x}^+x}\right\}e^{iq_{y}y},
\end{align}
\end{small}
where $q^{+}_{x}$ and $\pm q^{-}_{x}$ ($q^{\pm}_{x}=\sqrt{(E\mp M_L-V)^2/\hbar^2v_F^2 -q_y^2}$)
are, respectively, the momentum projections of the incident and spin-flip reflected electrons along the $x$-axis,
$\alpha^\pm=\arcsin(q_{y}/\sqrt{q_{x}^{\pm2}+q_{y}^2})$ are the corresponding incident and spin-flip reflected angles, and $r_{\downarrow\uparrow}$ and $r_{\uparrow\uparrow}$ are, respectively, the amplitudes of the spin-flip reflection and the normal reflection.
Here, we set $\tau=\pm1$ which means that the spin-flip reflected electron locates in the conduction band ($\tau=1$)
or the valence band ($\tau=-1$).
From the angle $\alpha^{-}$, we can find the outgoing direction of the spin-flip reflected electron.
If the incident electron and the spin-flip reflected electron locate in the same band,
the reflection is almost a specular reflection.
If they locate, respectively, in the conduction and valence bands, the reflection is almost a retro-reflection. As compared with the Andreev reflection in the graphene-superconductor interface,
this is exactly the opposite.
For the normal reflection, the reflected electron is always along the specular direction.

The wave function at the spin superconductor side ($x>0$) can be written as
\begin{small}
\begin{align}\label{eq:2e}
\psi(x,y)&=\left\{\frac{t_{\uparrow\uparrow}}{2\sqrt{\cosh\gamma}}\left(\begin{array}{c}
e^{-i\frac{\beta^+}{2}} e^{\frac{\gamma}{2}} \\
e^{i\frac{\beta^+}{2}} e^{\frac{\gamma}{2}} \\
e^{-i\frac{\beta^+}{2}} e^{-\frac{\gamma}{2}} \\
-e^{i\frac{\beta^+}{2}} e^{-\frac{\gamma}{2}} \\
\end{array}\right)e^{ik_{x}^+x}\right. \nonumber \\
+&\left.\frac{t_{\downarrow\uparrow}}{2\sqrt{\cosh\gamma}}\left(\begin{array}{c}
e^{i\frac{\beta^-}{2}}e^{-\frac{\gamma}{2}} \\
-e^{-i\frac{\beta^-}{2}}e^{-\frac{\gamma}{2}} \\
e^{i\frac{\beta^-}{2}}e^{\frac{\gamma}{2}} \\
e^{-i\frac{\beta^-}{2}}e^{\frac{\gamma}{2}} \\
\end{array}\right)e^{-ik_{x}^-x}\right\}e^{iq_{y}y},
\end{align}
\end{small}
where $k^{+}_{x}$ and $-k^{-}_{x}$ ($k^{\pm}_{x} = \sqrt{(-M\pm\sqrt{E^2-\Delta^2})^2/\hbar^2 v_F^2 -q_y^2} $)
are, respectively, the momentum projections of the normal and spin-flip transmitted electrons along the $x$-axis, $\beta^\pm=\arcsin(q_{y}/\sqrt{k_{x}^{\pm2}+q_{y}^2})$ are the refraction angles, and
$e^{\gamma}\equiv(E+\sqrt{E^2-\Delta^2})/\Delta$. $t_{\downarrow\uparrow}$ and $t_{\uparrow\uparrow}$ are the amplitudes of the spin-flip transmission and the normal transmission, respectively.

After obtaining the wave functions at the $x<0$ and $x>0$ sides, the next procedure is to connect them at the
interface $x=0$.
For the linear dispersion case, the wave-function should be continuous at the interface, i.e.,
$\psi(0^-,y)=\psi(0^+,y)$. By substituting Eqs.(\ref{eq:2d}) and (\ref{eq:2e}) into the boundary condition,
the reflection and transmission amplitudes can be obtained straightforwardly:
\begin{small}
\begin{align}\label{eq:2f}\left\{\begin{array}{ll}
r_{\downarrow\uparrow}&=\cos\alpha^+\cos\left(\frac{\beta^++\beta^-}{2}\right)/\Gamma  \\
r_{\uparrow\uparrow}&=i\left\{e^\gamma\cos\left(\frac{\tau\alpha^--\beta^-}{2}\right)\sin\left(\frac{\alpha^+-\beta^+}{2}\right)\right. \\
&\left.+e^{-\gamma}\sin\left(\frac{\tau\alpha^-+\beta^+}{2}\right)\cos\left(\frac{\alpha^++\beta^-}{2}\right)\right\}/\Gamma \\
t_{\uparrow\uparrow}&=e^{\frac{\gamma}{2}}\cos\alpha^+\cos\left(\frac{\tau\alpha^--\beta^-}{2}\right)\sqrt{2\cosh\gamma}/\Gamma  \\
t_{\downarrow\uparrow}&=ie^{-\frac{\gamma}{2}}\cos\alpha^+\sin\left(\frac{\tau\alpha^-+\beta^+}{2}\right)\sqrt{2\cosh\gamma}/\Gamma \end{array}\right.
\end{align}
\end{small}
where
\begin{small}
\begin{align}\label{eq:2g}
\Gamma=&e^\gamma\cos\left(\frac{\tau\alpha^--\beta^-}{2}\right)\cos\left(\frac{\alpha^++\beta^+}{2}\right) \nonumber \\
&+e^{-\gamma}\sin\left(\frac{\tau\alpha^-+\beta^+}{2}\right)\sin\left(\frac{\alpha^+-\beta^-}{2}\right).
\end{align}
\end{small}

\begin{figure}[h]
\includegraphics[width=1.0\columnwidth]{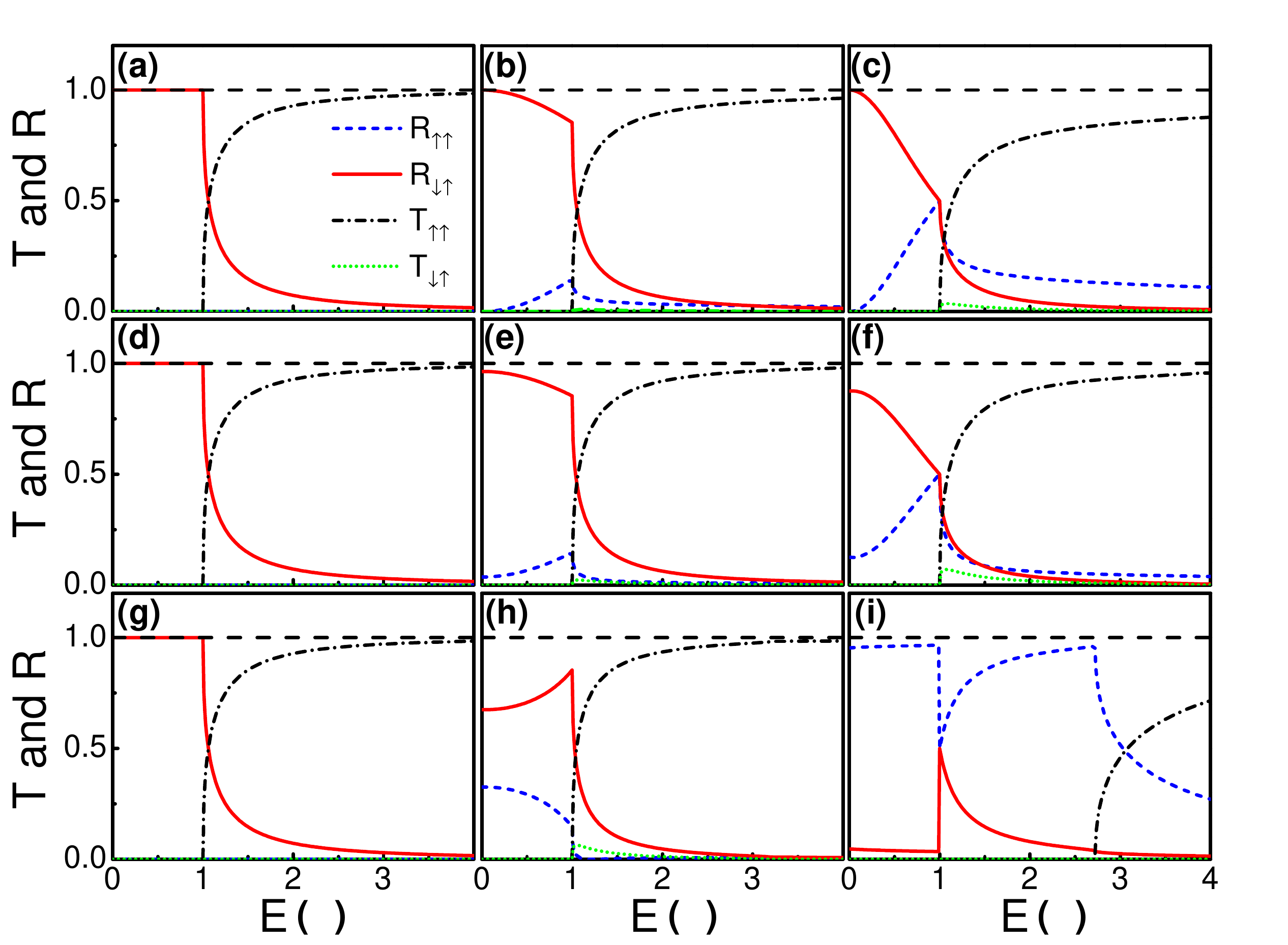}
\caption{(Color online) Transmission and reflection coefficients of a spin-up electron incident
from the left side with $M_L=0$, $\Delta=1$, and $M=-10\Delta$.
The incident angle $\alpha^{+}$ is $0$, $\pi/8$, and $\pi/4$ for the first column [(a), (d), and (g)], for
the second column [(b), (e), and (h)], and for the third column [(c), (f), and (i)], respectively.
The potential energy is $V=0$ for the first row (a)-(c), $V=-5\Delta$ for the second row (d)-(f), and $V=-15\Delta$
for the third row (g)-(i).
}  \label{fig2}
\end{figure}

The corresponding transmission and reflection coefficients can be derived by using the conservation law
$\nabla \cdot \bm{j}=-{\frac{\partial}{ \partial t}} |\psi|^2$.
Here, the probability current $\bm{j}$ is given by
\begin{align}\label{eq:g1}
\bm{j}=v_{F}(\psi^\dagger\sigma_{0}\otimes\tau_{x}\psi \bm{e_{x}}+\psi^\dagger\sigma_{0}\otimes\tau_{y}\psi \bm{e_{y}} ),
\end{align}
where $\sigma_{0}$ is the $2\times2$ identity matrix in spin space and $\bm{\tau}\equiv(\tau_{x}, \tau_{y})$ is the 2D vector of Pauli
matrices in pseudospin space.
Substituting Eqs.(\ref{eq:2d}) and (\ref{eq:2e}) into Eq.(\ref{eq:g1}), we obtain the relation between the scattering coefficients and the scattering amplitudes as
\begin{align}\label{eq:2h}
\left\{\begin{array}{l}
R_{\downarrow\uparrow}=|r_{\downarrow\uparrow}|^2 \frac{\cos{\alpha^-}}{\cos{\alpha^+}} , \\
R_{\uparrow\uparrow}=|r_{\uparrow\uparrow}|^2  ,  \\
T_{\uparrow\uparrow}=|t_{\uparrow\uparrow}|^2 \frac{\cos{\bm{Re}\beta^+}}{\cos{\alpha^+}}\frac{\sinh{\bm{Re}{\gamma}}}{|\cosh{\gamma}|} ,\\
T_{\downarrow\uparrow}=|t_{\downarrow\uparrow}|^2\frac{\cos{\bm{Re}\beta^-}}{\cos{\alpha^+}}\frac{\sinh{\bm{Re}{\gamma}}}{|\cosh{\gamma}|}.
\end{array}\right.
\end{align}
Here, $T_{\uparrow\uparrow}+T_{\downarrow\uparrow}+R_{\uparrow\uparrow}+R_{\downarrow\uparrow}=1$ because of the
current conservation. One can see
from Eq.(\ref{eq:2h}) that $T_{\uparrow\uparrow}=T_{\downarrow\uparrow}=0$ exactly
when the energy $|E|$ of the incident electron is less than $\Delta$, because of the existence of the gap of the spin superconductor. Then, $R_{\uparrow\uparrow}+R_{\downarrow\uparrow}=1$ for $|E|<\Delta$.

First of all, we present the transmission and reflection coefficients of the graphene-based
normal-spin superconductor junction with the magnetic momentum at the left side being $M_{L}=0$, as shown in Fig.\ref{fig2}.
If an electron is injected normally to the interface ($\alpha^+=0$),
the spin-flip reflection is perfect with $R_{\downarrow\uparrow}=1$
and the normal reflection completely vanishes for $|E|<\Delta$,
regardless of the potential $V$ [see Figs.\ref{fig2}(a), \ref{fig2}(d), and \ref{fig2}(g)].
The spin-flip reflection can be slightly suppressed by either the oblique incidence or the decrease of the potential $V$.
Here, we emphasize that for $|E|<\Delta$, the spin-flip process dominates the reflection phenomenon
and $R_{\downarrow\uparrow}>R_{\uparrow\uparrow}$ holds for a wide range of parameters,
except for $\alpha^{+}$=$\pi/4$ and $V=-15\Delta$ in Fig.\ref{fig2}(i).
These results can be explained as follows. The reflected angle and the refraction one satisfy the relation $\sin{\alpha^-}=\frac{E-M_{L}-V}{E+M_{L}-V}\sin{\alpha^+}$
and $\sin{\beta^\pm}=\frac{E-M_{L}-V}{-M\pm\sqrt{E^2-\Delta^2}}\sin{\alpha^+}$. Normal incidence ($\alpha^+=0$) with $M_{L}=0$ indicates $\alpha^+=\alpha^-=\beta^+=\beta^-=0$,
leading to $R_{\downarrow\uparrow}=|e^{-2\gamma}|$ which is 1 for $|E|<\Delta$ and decays quadratically for $E>\Delta$.
Both terms of $\cos{\alpha^+}$ and
$\cos{\frac{\beta^++\beta^-}{2}}$ in $r_{\downarrow\uparrow}$ will be reduced by increasing the incident angle $\alpha^+$ [see Eq.(17)], which slightly weakens the spin-flip effects.
From a physical viewpoint, when an electron is obliquely incident, $q_{y}$ is a conserved quantity and the dispersion spectrum of $q_{x}$ is no longer linear. Then,
the Fermi wave lengths mismatch suppresses the spin-flip reflection.
Note that the normal transmission appears immediately
when the incident energy $E$ exceeds the energy gap $\Delta$ in general.
However, one can see from Fig.\ref{fig2}(i) that the normal transmission appears only when $E$ exceeds
the critical energy $E_{c}\approx2.72\Delta$.
The underlying physics is that there does not exist any corresponding scattering state in the spin superconductor for an obliquely incident electron, and the critical condition is determined by the relation $\sqrt{E^2-\Delta^2}-M\geq(E-V)\sin{\alpha^+}$. By solving the above equation, the critical energy can be obtained as
$E_{c1}=-D/\sin{\alpha^+}$ and $E_{c2}=\frac{\sin{\alpha^+}D+\sqrt{D^2+\Delta^2\cos^2 {\alpha^+}}}{\cos^2{\alpha^+}}$, where $D=M-V\sin{\alpha^+}$.
The prerequisite for the appearance of the normal transmission is $\Delta\leq E\leq E_{c1}$ or $E\geq \max(E_{c2}, \Delta)$. In Fig.\ref{fig2}(i), the critical energy is $E_{c2}\approx2.72\Delta$ because $E_{c1}<0$.

\begin{figure}[h]
\includegraphics[width=1.0\columnwidth]{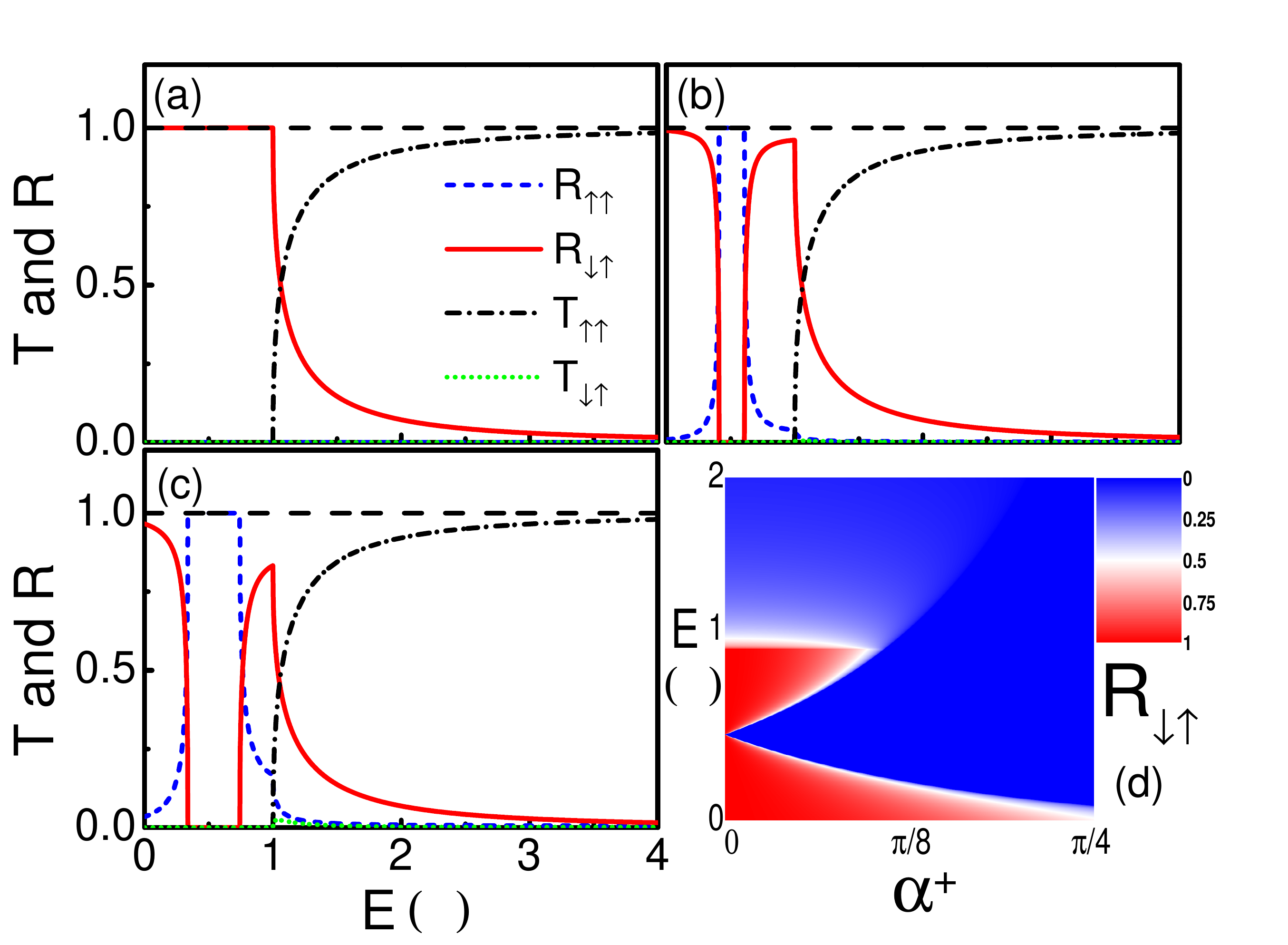}
\caption{(Color online) (a-c) Transmission and reflection coefficients of the graphene-based normal-spin superconductor junction as a function of the incident energy $E$, with (a) $\alpha^{+}=0$, (b) $\alpha^{+}=\pi/32$, and (c) $\alpha^{+}=\pi/16$. The other parameters are $\Delta=1$, $M_{L}=-0.5\Delta$, $M=-10\Delta$, and $V=0$. (d) A 2D plot of spin-flip reflection coefficient $R_{\downarrow\uparrow}$
versus $\alpha^{+}$ and $E$. The parameters are the same as (a-c).} \label{fig3}
\end{figure}

Next, we study the influence of nonzero magnetic momentum $M_{L}$ on the scattering coefficients.
Fig.\ref{fig3} shows the transmission and reflection coefficients with $M_{L}=-0.5\Delta$ and the potential $V=0$.
For the normal incidence ($\alpha^+=0$), the nonzero $M_{L}$ will not affect the scattering coefficients as compared with Fig.\ref{fig2}(a), and the spin-flip reflection coefficient remains $R_{\downarrow\uparrow}=1$ for $|E|<\Delta$ [see Fig.\ref{fig3}(a)].
The oblique incidence with nonzero $\alpha^+$ induces a region
in which the spin-flip reflection coefficient is $R_{\downarrow\uparrow}=0$, due to the absence of the spin-flip scattering state in this region.
The larger the angle $\alpha^+$ is, the wider region of $R_{\downarrow\uparrow}=0$, as illustrated in Figs.\ref{fig3}(b) and \ref{fig3}(c). While out of this region, $R_{\downarrow\uparrow}$ is very large.
Here, the two regions of large $R_{\downarrow\uparrow}$ denote the spin-flip specular reflection
and the spin-flip retro-reflection.
In the region of $|E|<|M_L|$, the incident electron and the spin-flip reflected electron locate, respectively, in
the conduction band and the valence one, and the spin-flip retro-reflection appears.
In the region of $|E|>|M_L|$, the incident and spin-flip reflected electrons locate in
the same band, and the spin-flip specular reflection occurs.
In Fig.\ref{fig3}(d), the spin-flip reflection coefficient $R_{\downarrow\uparrow}$ is plotted as functions of $\alpha^{+}$ and $E$, with the incident angle $\alpha^{+}$ changing from $0$ to $\pi/4$.
There are two regions in which $R_{\downarrow\uparrow}$ is almost 1.
These two regions are the spin-flip specular reflection and the spin-flip retro-reflection.

\begin{figure}[h]
\includegraphics[width=1.0\columnwidth]{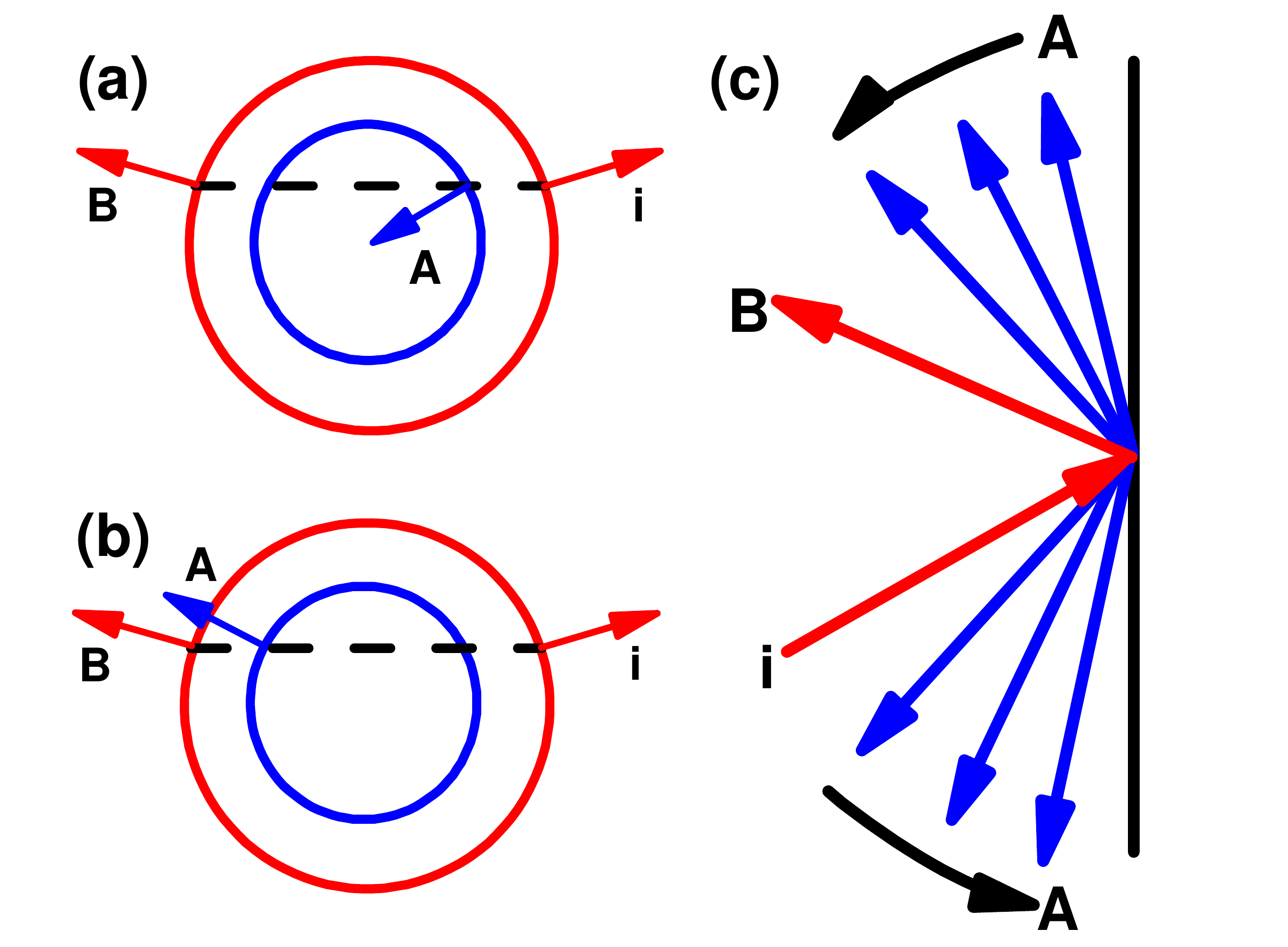}
\caption{(Color online) (a) and (b) Cross sections of the energy spectrum of the graphene
with $M_L\not=0$
in Fig.\ref{fig1}. The red lines denote the dispersion spectrum of the spin-up states
and the blue lines represent the spin-down ones.
The direction of motion of the states is indicated by the arrows.
A spin-up electron in the conduction band incident from the graphene
can be either spin-flip retro-reflected (a) or spin-flip specular reflected (b),
depending on whether the reflected electron locates in the valence band (a) or the conduction band (b).
(c) Trajectories of an incident electron i, the normal reflected electron B, and
the spin-flip reflected electron A, for different incident energies $E$ by fixing the incident
angle.
By increasing the incident energy $E$ from $0$, although the normal reflected angle remains the same,
the spin-flip angle rotates anticlockwise that it disappears and reappears when the spin-flip reflected
state is shifted from the valence band to the conduction one.
} \label{fig4}
\end{figure}

This peculiar reflection property can be well demonstrated using the schematic diagram illustrated in Fig.\ref{fig4}. Figs.\ref{fig4}(a) and \ref{fig4}(b) present the cross sections of the energy spectrum of the graphene with nonzero $M_L$, which are obtained by cutting along the dotted line in Fig.\ref{fig1}.
A spin-up electron $i$ incident from the conduction band of the graphene can be either spin-flip reflected $A$ or normal reflected $B$ at the normal-spin superconductor interface.
One can see from Figs.\ref{fig4}(a) and \ref{fig4}(b) that the normal reflection is always specular,
whereas the spin-flip reflection can be either retro (see Fig.\ref{fig4}(a)) or specular (see Fig.\ref{fig4}(b)), depending on whether the reflected electron locates in the valence band or the conduction band of the spin-down energy spectrum.
For a certain oblique angle $\alpha^+$, when the incident energy $E$ is increased from $0$, the normal reflected angle remains constant and the spin-flip reflected angle $\alpha^-$ rotates anticlockwise that it disappears for a while, then reappears, and moves closer to the normal reflected angle [see Fig.\ref{fig4}(c)].
The critical energy for the disappearance or the reappearance of the spin-flip reflection can be obtained as follows: the spin-flip reflected angle $\alpha^-$ and the incident angle $\alpha^+$ satisfy the relation
$\sin{\alpha^-}=\frac{E-M_{L}}{\tau(E+M_{L})}\sin{\alpha^+}$.
The condition of $\sin{\alpha^-}\leq1$ gives two critical energies $E_{c,\pm}=-M_{L}\frac{1\pm\sin{\alpha^+}}{1\mp\sin{\alpha^+}}$, between which
the spin-flip reflection disappears.

\begin{figure}[h]
\includegraphics[width=1.0\columnwidth]{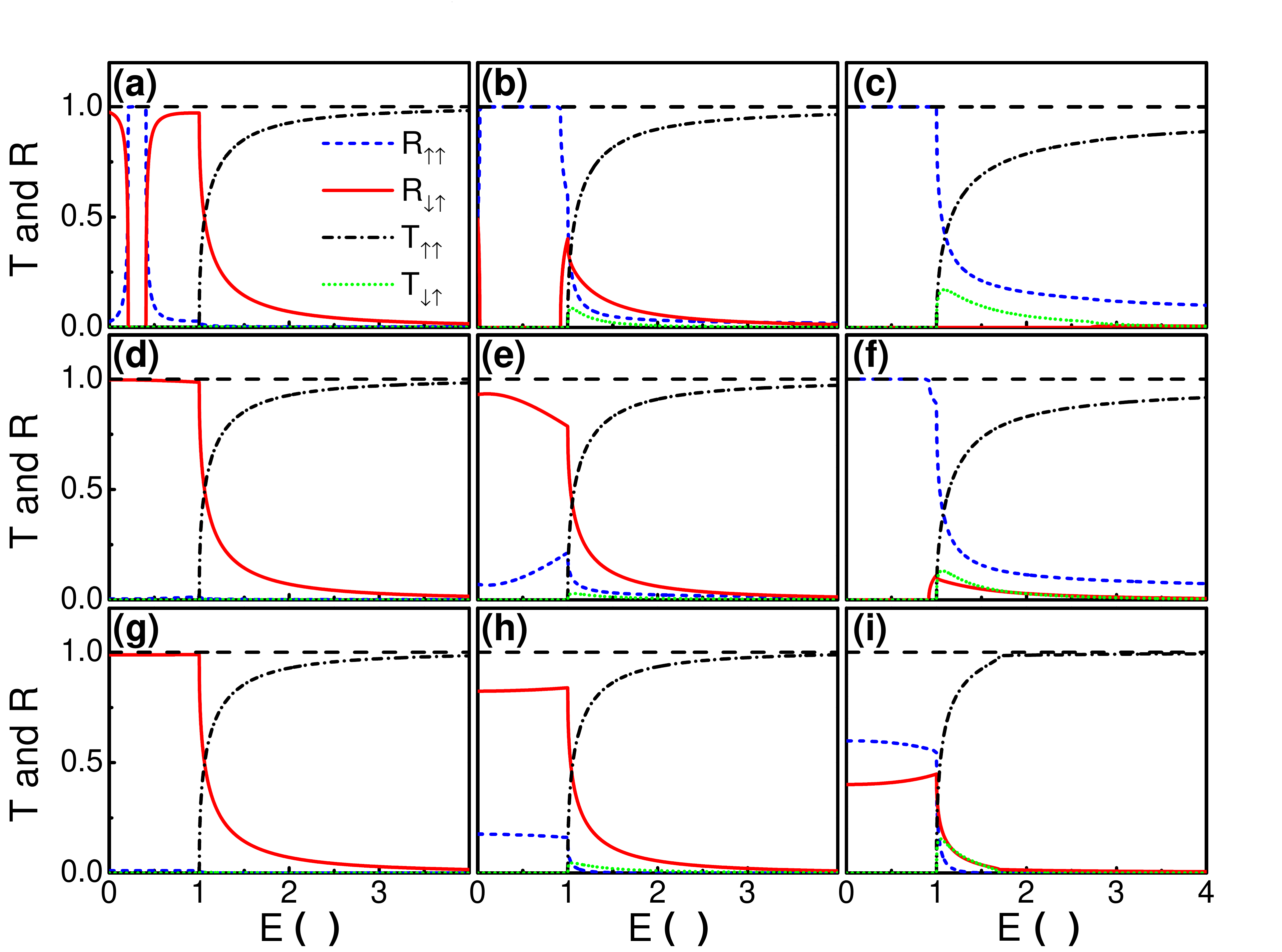}
\caption{(Color online) Transmission and reflection coefficients of a spin-up electron incident
from the left side with $\Delta=1$, $M_{L}=-0.5\Delta$, and $M=-10\Delta$.
The incident angle $\alpha^{+}$ is $\pi/32$, $\pi/8$, and $\pi/4$ for the first column [(a), (d), and (g)], for
the second column [(b), (e), and (h)], and for the third column [(c), (f), and (i)], respectively.
The potential energy is $V=-0.2\Delta$ for the first row (a)-(c), $V=-2\Delta$ for the second row (d)-(f), and $V=-10\Delta$ for the third row (g)-(i).}
\label{fig5}
\end{figure}

Fig.\ref{fig5} shows the transmission and reflection coefficients in the case of nonzero $M_L$ and nonzero potential $V$.
For small incident angles $\alpha^+$, e.g., $\alpha^+=\pi/32$, when the potential
$V$ is decreased from $0$ to $-10\Delta$ which can be experimentally realized by the gate voltage,
the region of the spin-flip retro-reflection diminishes [see Figs.\ref{fig3}(b) and \ref{fig5}(a)]
and disappears for $V<-\Delta$ [see Figs.\ref{fig5}(d) and \ref{fig5}(g)], whereas the region of the spin-flip specular reflection is enlarged. For $V < -\Delta$ (see, e.g., Figs.\ref{fig5}(d) and \ref{fig5}(g)),
the spin-flip specular reflection
occurs almost perfectly with the spin-flip reflection coefficient $R_{\downarrow\uparrow}\approx 1$ for $|E|<\Delta$.
On the other hand, for relative large incident angles $\alpha^+$, e.g., $\alpha^+=\pi/8$ and $\pi/4$, the spin-flip reflection coefficient $R_{\downarrow\uparrow}$ is usually small and the normal reflection coefficient $R_{\uparrow\uparrow}$ is large for small $|V|$ (see Figs.\ref{fig5}(b), \ref{fig5}(c), and \ref{fig5}(f)). By decreasing the potential $V$, the spin-flip reflection
is considerably strengthened and the normal reflection is weakened (see Figs.\ref{fig5}(e), \ref{fig5}(h), and \ref{fig5}(i)).
The larger the incident angle $\alpha^+$ is, the larger $|V|$ is needed to strength the spin-flip reflection.
In fact, by decreasing $V$, the critical energy $E_{c,+}$ is reduced. When $E_{c,+}<\Delta$, the spin-flip reflection is dramatically enhanced and dominates the reflection process.

It is known that when a voltage is applied between a normal metal and a charge superconductor, an electric current can be generated. This is due to the occurrence of the Andreev reflection at the normal metal-superconductor interface when a charge is injected into the superconductor.
The quasi-particle current in the normal metal will be transformed to the supercurrent in the superconductor. However,
at a normal metal-spin superconductor interface, it is the spin rather than the charge degree of freedom that plays a role.
An electron will be spin-flip reflected as another electron, with spin angular momentum $2\times\frac{\hbar}{2}$ being injected into the spin superconductor.
Therefore, the quasi-particle spin current in the normal metal will be transformed to a super spin current
in the spin superconductor.
Here, we consider both the charge voltage and the spin voltage.
Under the voltage, the chemical potential of the spin superconductor is fixed to be $\mu_{R}=0$,
and the chemical potential of the normal metal is set to be $\mu_{L\uparrow}=\mu_{L\downarrow}=eV_{s}$ for the charge voltage and $\mu_{L\uparrow}=-\mu_{L\downarrow}=eV_{s}$ for the spin voltage.\cite{addr4,addr5}
The spin voltage can be experimentally implemented by various methods with today's technologies, for example by using the spin Hall effect,\cite{sv11,sv12,sv13}
the spin Seebeck effect,\cite{sv1,sv2,sv3} and the optical excitation.\cite{newnew}
After obtaining the transmission and reflection coefficients, the particle current $I_{\uparrow/\downarrow}$ can be calculated from the Landauer-B\"{u}ttiker formula,\cite{sca3}
\begin{eqnarray}
I_{\sigma} &= &\frac{1}{S} \sum\limits_{k_x,k_y}\left[ (R_{\bar{\sigma}\sigma}+T_{\sigma\sigma}+T_{\bar{\sigma}\sigma})v_x f_{L\sigma} -
R_{\sigma\bar{\sigma}}v_x f_{L\bar{\sigma}}\right], \nonumber
\end{eqnarray}
where $\bar{\sigma}=\downarrow (\uparrow)$ when $\sigma=\uparrow (\downarrow)$,
$f_{L\sigma}(E)$ is the Fermi distribution function with $f_{L\sigma}(E)=\Theta(\mu_{L\sigma}-E)$ at zero
temperature, $S$ is the total area of the graphene sheet, and
$v_{x}$ is the electron velocity along the positive $x$-axis.
Here, the sum is performed over the states with positive $v_x$.
Then, the spin and charge currents can be obtained as $I_{s}=\frac{\hbar}{2}(I_{\uparrow}-I_{\downarrow})$ and $I_{e}=e(I_{\uparrow}+I_{\downarrow})$.
And the corresponding differential conductance is $G_{e,s}=\frac{dI_{e,s}}{dV_{s}}$.

\begin{figure}[h]
\includegraphics[width=1.0\columnwidth]{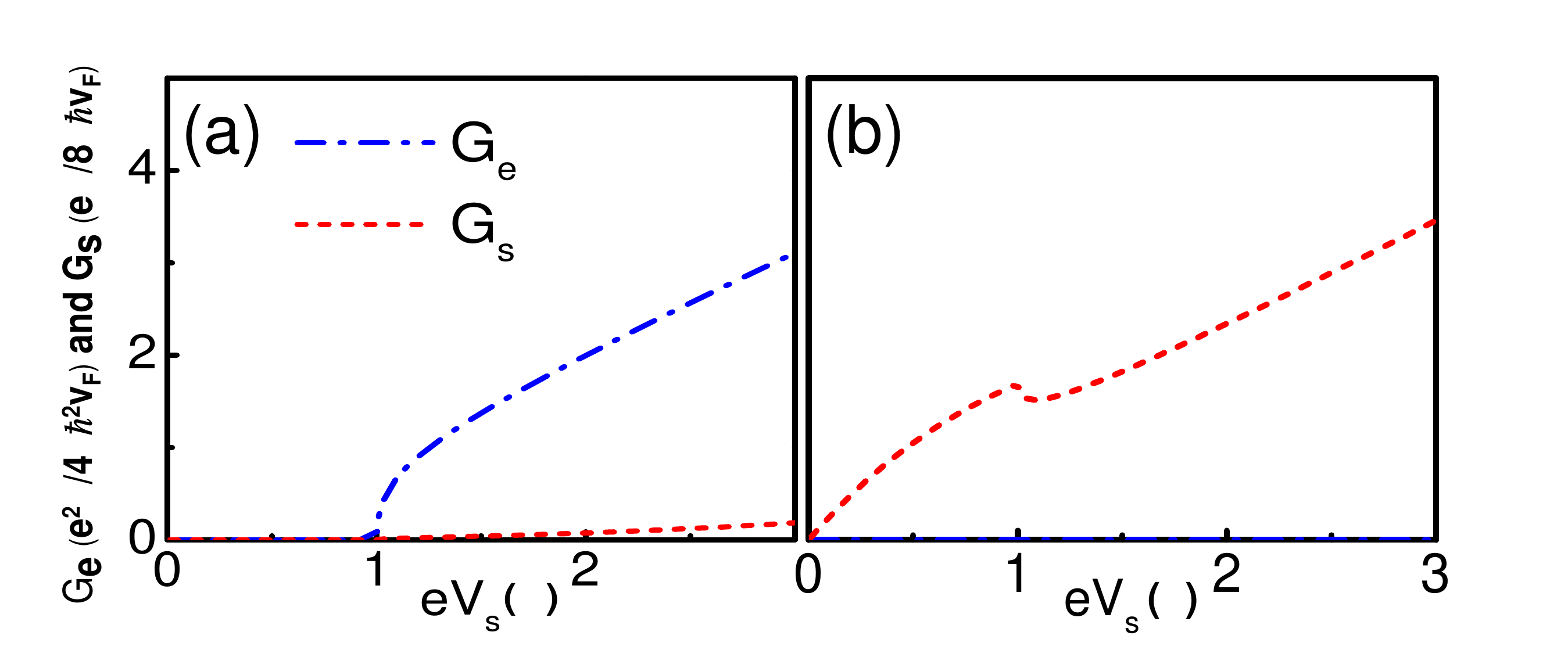}
\caption{(Color online) Differential charge conductance $G_{e}$ and differential spin conductance $G_{s}$
versus the charge voltage $eV_{s}$ (a) and versus the spin voltage $eV_{s}$ (b). The parameters are $M_{L}=0$, $V=0$, $\Delta=1$, and $M=-10\Delta$.} \label{fig6}
\end{figure}

Fig.\ref{fig6} displays the differential spin conductance $G_{s}$ and the charge conductance $G_{e}$ under the charge voltage and the spin voltage.
One can see that in the case of the charge voltage, both the spin conductance $G_{s}$ and the charge conductance $G_{e}$ are zero when the voltage is smaller than the energy gap $\Delta$ [see Fig.\ref{fig6}(a)],
because of the gap in the spin superconductor and the spin superconductor is a charge insulator.
When the voltage $eV_{s}$ is larger than $\Delta$, both $G_{s}$ and $G_{e}$ increase linearly with $eV_{s}$, owing to the fact that the quasi-particle states locate above the gap of the spin superconductor and the linear dispersion relation of graphene.
On the other hand, in the case of the spin voltage, although the voltage satisfies $eV_{s}<\Delta$,
$G_{s}$ grows quickly with the voltage $eV_{s}$, whereas $G_{e}$ is strictly zero [see Fig.\ref{fig6}(b)], due to the occurrence of the spin-flip reflection.
In this situation, the quasi-particle spin current in the normal metal flows through the interface and is transformed to a super spin current when it enters into the spin superconductor.
Although the charge current is zero, a large spin current can flow through the junction even if the spin voltage $eV_{s}$ is small.
This means that the normal metal-spin superconductor junction is transparent for the spin current,
owing to the contribution of the spin-flip reflection.
When the voltage satisfies $eV_{s}>\Delta$, $G_{s}$ is still large and increases linearly with $eV_{s}$ because of the tunneling of the quasi-particle ($T_{\downarrow\uparrow}$ and $T_{\uparrow\uparrow}$). However, $G_e$ is exactly zero because of the particle-hole symmetrical band structure at $M_L =0$ and $V=0$, and remains very small when $M_L \not=0$ and $V\not=0$.

\section{\label{sec:quadratic}Quadratic dispersion case}

In this section, we calculate the 1D reflection and the transmission
coefficient of the normal metal-spin superconductor junction, where both the normal metal and the spin superconductor have quadratic dispersion.
The Hamiltonian corresponding to this case is shown in Eq.(\ref{eq:1k}).
The wavefunctions on both sides are assumed to be plane waves $\psi(x)=(a,b)^Te^{iqx}$.
There are two degenerate energy bands at the normal side
\begin{align}\label{eq:3a}
E_{N}(q)=\frac{\hbar^2q^2}{2m}+V.
\end{align}
For a certain incident energy $E>0$, there are four degenerate states
$\psi_{N,\uparrow}^{\pm}(x)=(1,0)^T\exp(\pm iqx)$
and $\psi_{N,\downarrow}^{\pm}(x)=(0,1)^T\exp(\pm iqx)$,
where $q=\sqrt{\frac{2m(E-V)}{\hbar^2}}$.
The indices $\pm$ in $\psi_{N,\uparrow(\downarrow)}^{\pm}$ denote the velocity direction of the states. For example, $\psi_{N,\uparrow}^{+}$ describes a spin-up electron which moves along the positive direction (from the left to the right).

\begin{figure}[h]
\includegraphics[width=1.0\columnwidth]{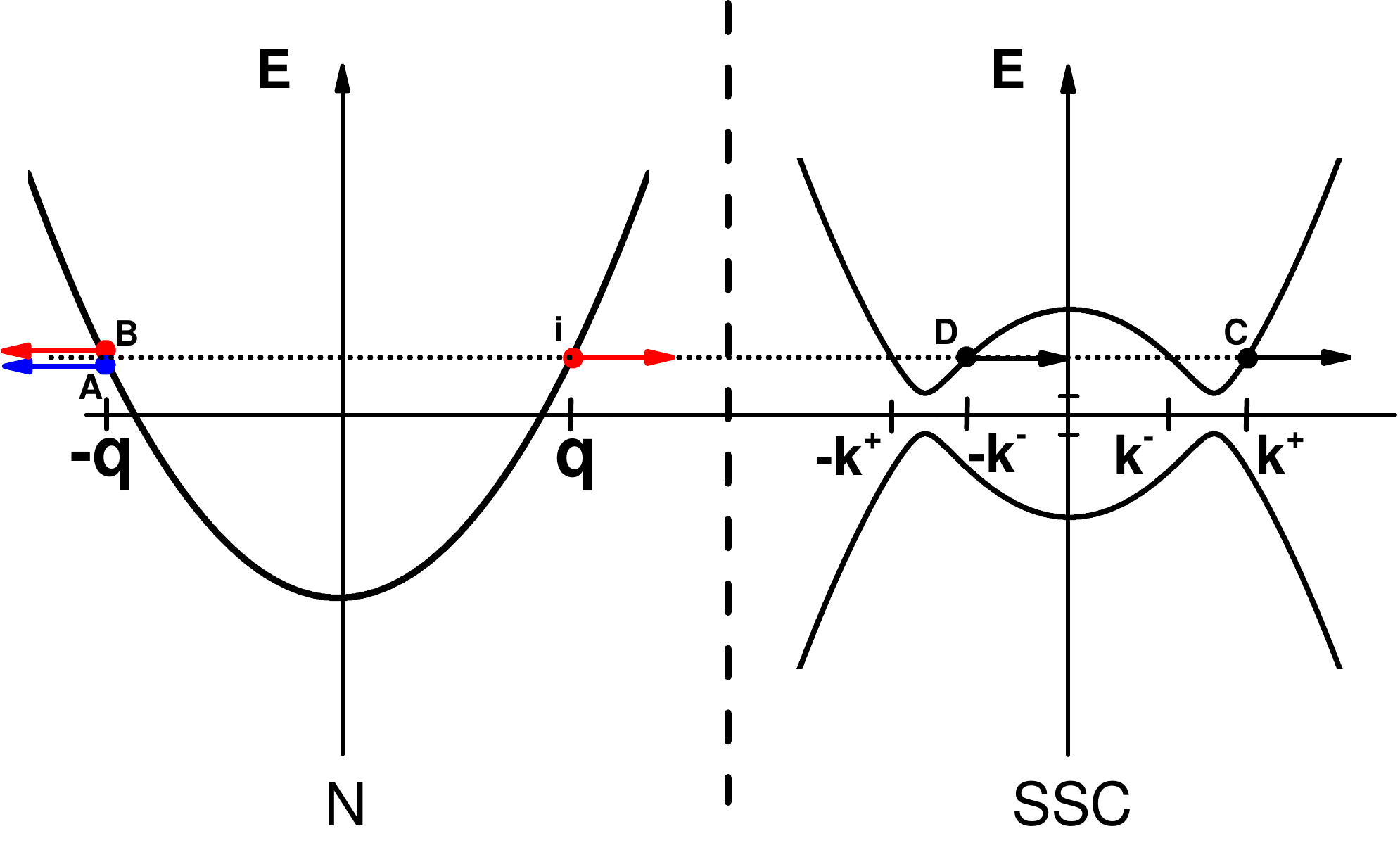}
\caption{(Color online) Band structures of the normal metal (N) and the spin superconductor (SSC) for the quadratic dispersion case. In the normal metal, the spin degree of freedom is degenerate, and the spin-up states are denoted by the red lines and the spin-down states are denoted by the blue line. A spin-up electron i with wave vector $q$ is incident from the normal metal to the spin superconductor. Spin-flip reflection A, normal reflection B, normal transmission C, and spin-flip transmission D are indicated by the arrows.} \label{fig7}
\end{figure}

In the spin superconductor, the energy bands are
\begin{align}\label{eq:3d}
E_{S}(k)=\pm\sqrt{\left(\frac{\hbar^2k^2}{2m}+M\right)^2+\Delta^2}.
\end{align}
The four corresponding degenerate states are
\begin{align}\label{eq:3e}
\psi_{S,\uparrow}^{\pm}(x)=\left( \begin{array}{c}
u_0 \\
v_0 \\
\end{array}\right)\exp{\left[\pm ik^{+}x\right]}
\end{align}
and
\begin{align}\label{eq:3f}
\psi_{S,\downarrow}^{\pm}(x)=\left( \begin{array}{c}
v_0 \\
u_0 \\
\end{array}\right)\exp{\left[\mp ik^{-}x\right]},
\end{align}
where $k^{\pm}(E)=\sqrt{\frac{2m(-M\pm\sqrt{E^2-\Delta^2})}{\hbar^2}}$,
$u_{0}^2=\frac{1}{2}(1+\frac{\sqrt{E^2-\Delta^2}}{E})$, and $v_{0}^2=\frac{1}{2}(1-\frac{\sqrt{E^2-\Delta^2}}{E})$.
$u_{0}$ and $v_{0}$ are the coherent factors of the spin superconductor, which also appear in the case of the
Bogoliubov-de Gennes (BdG) equation of the charge superconductor. The band structures of the normal metal and the spin superconductor are illustrated in Fig.\ref{fig7}.

We consider the situation where a spin-up electron is incident from the normal metal to the interface. By connecting the wave functions in both regions at $x=0$, the reflection and transmission coefficients can be obtained.
By using the equations derived above, the wave functions in the normal metal can be expressed as
\begin{small}
\begin{align}\label{eq:3g}
\psi_{1}(x)=\left(\begin{array}{c}
1 \\
0 \\
\end{array}\right)e^{iqx}+r_{\downarrow\uparrow}\left(\begin{array}{c}
0 \\
1 \\
\end{array}\right)e^{-iqx}+r_{\uparrow\uparrow}\left(\begin{array}{c}
1 \\
0 \\
\end{array}\right)e^{-iqx}.
\end{align}
\end{small}
In the spin superconductor, the wave functions are
\begin{small}
\begin{align}\label{eq:3h}
\psi_{2}(x)=t_{\uparrow\uparrow}\left(\begin{array}{c}
u_{0} \\
v_{0} \\
\end{array}\right)e^{ik^{+}x}+t_{\downarrow\uparrow}\left(\begin{array}{c}
v_{0} \\
u_{0} \\
\end{array}\right)e^{-ik^{-}x}.
\end{align}
\end{small}
The boundary conditions of the wave functions at $x=0$ are
\begin{align}\label{eq:3i}
\left\{\begin{array}{l}
\psi(0^-)=\psi(0^+) ,\\
\sigma^z\psi'(0^+)-\psi'(0^-)=\frac{2mV_0}{\hbar^2}\psi(0) .\\
\end{array}\right.
\end{align}
The first line comes from the continuity of the wave functions and the second one is the requirement of the $\delta$ potential.

\begin{figure}[h]
\includegraphics[width=1.0\columnwidth]{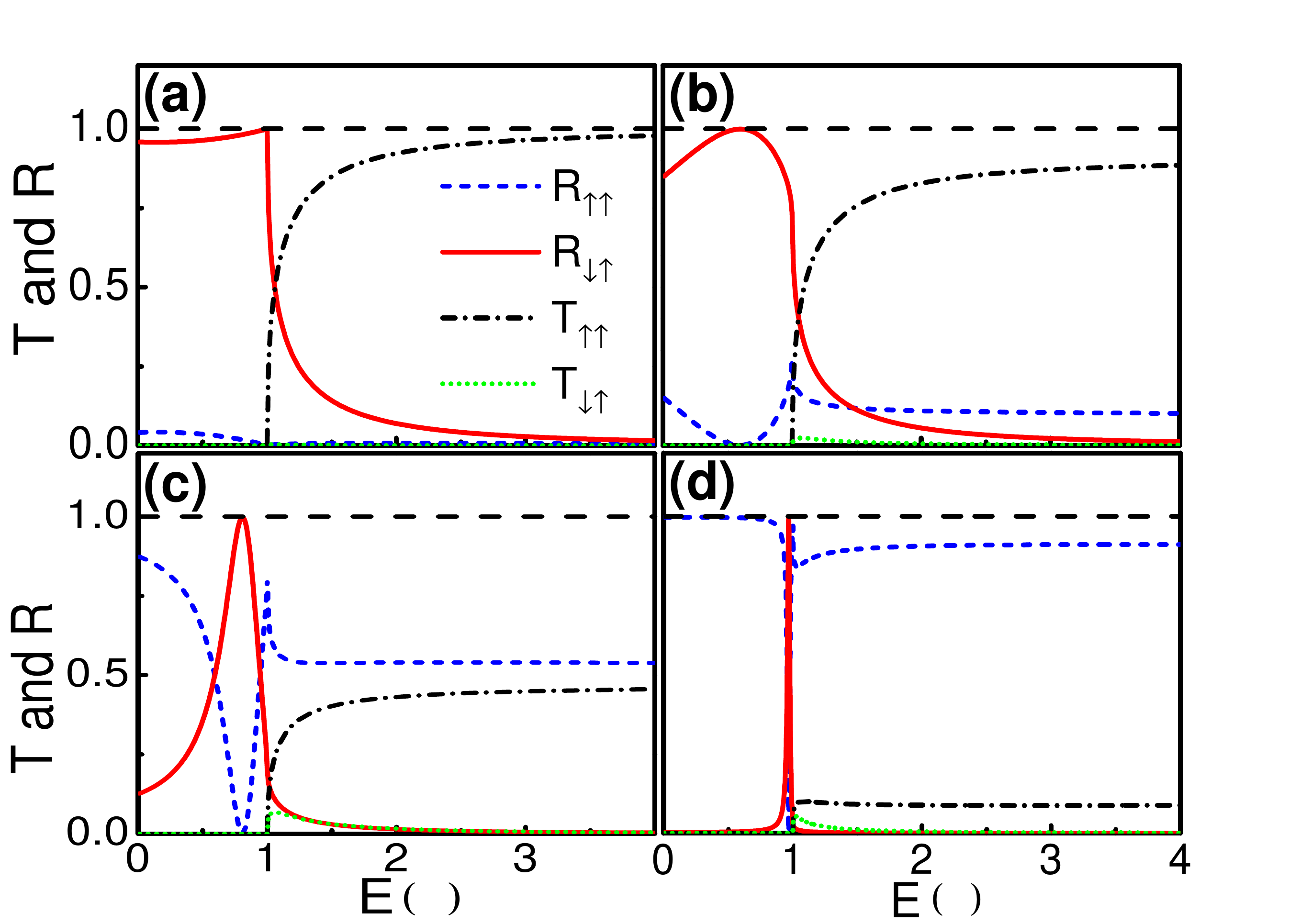}
\caption{(Color online) Transmission and reflection coefficients versus the incident energy $E$
for the normal metal-spin superconductor junction with the quadratic dispersion case.
The Schottky potential strength is $Z=0$ (a), $0.3$ (b), $1.0$ (c), and $3.0$ (d).
The other parameters are $\Delta=1$, $V=-15\Delta$, and $M=-10\Delta$.} \label{fig8}
\end{figure}

Eq.(\ref{eq:3i}) determines the four parameters $r_{\downarrow\uparrow}$, $r_{\uparrow\uparrow}$, $t_{\uparrow\uparrow}$, and $t_{\downarrow\uparrow}$ uniquely. After some analytical calculations,  the reflection and transmission amplitudes can be derived straightforwardly
\begin{align}\label{eq:3j}\left\{\begin{array}{l}
r_{\downarrow\uparrow}=\frac{2u_0v_0[\eta^{+}+\eta^{-}]}{\Gamma}  \\
r_{\uparrow\uparrow}=-1+\frac{2u_0^{2}(1+2iZ+\eta^{-})-2v_0^{2}(1+2iZ-\eta^{+})}{\Gamma} \\
t_{\uparrow\uparrow}=\frac{2u_0[1+2iZ+\eta^{-}]}{\Gamma}  \\
t_{\downarrow\uparrow}=-\frac{2v_0[1+2iZ-\eta^{+}]}{\Gamma}
\end{array}\right.
\end{align}
where $\eta^{\pm}=k^{\pm}/q$ are dimensionless quantities, $Z=mV_{0}/\hbar^2q$ measures the barrier strength, and
\begin{align}\label{eq:3k}
\Gamma=&u_0^2[1+2iZ+\eta^{-}][1+2iZ+\eta^{+}] \nonumber \\
-&v_0^2[1+2iZ-\eta^{-}][1+2iZ-\eta^{+}].
\end{align}

By using the conservation of probability $\nabla \cdot \bm{j}=-{\frac{\partial}{ \partial t}} |\psi|^2$, the probability current is then given by
$j_{x}=\frac{\hbar}{m}\bm{Im}(\psi^{\dag}\sigma_{0}\partial_{x}\psi)$ at $x<0$ and $j_{x}=\frac{\hbar}{m}\bm{Im}(\psi^{\dag}\sigma_{z}\partial_{x}\psi)$ at $x>0$.
By substituting Eqs.(\ref{eq:3g}) and (\ref{eq:3h}) into the expression of the probability current,
the relation between the transmission (reflection) coefficients and the amplitudes can be obtained as
\begin{align}\label{eq:3l}\left\{\begin{array}{l}
R_{\uparrow\uparrow}=|r_{\uparrow\uparrow}|^2 , \\
R_{\downarrow\uparrow}=|r_{\downarrow\uparrow}|^2 ,\\
T_{\uparrow\uparrow}=(|u_0|^2-|v_0|^2)\eta_{+}|t_{\uparrow\uparrow}|^{2} ,\\
T_{\downarrow\uparrow}=(|u_0|^2-|v_0|^2)\eta_{-}|t_{\downarrow\uparrow}|^{2} .
\end{array}\right.
\end{align}
Here, the four coefficients satisfy $R_{\uparrow\uparrow}+R_{\downarrow\uparrow} +T_{\uparrow\uparrow}+T_{\downarrow\uparrow} =1$, due to the current conservation.

Fig.\ref{fig8} plots the reflection coefficients $R$ and the transmission coefficients $T$ for various barrier strengths $Z$.
When the incident energy $E$ is less than the gap $\Delta$, the transmission coefficients $T_{\uparrow\uparrow}$ and $T_{\downarrow\uparrow}$ are zero exactly, due to the absence of the state in the spin superconductor when $|E|<\Delta$. As a result, $R_{\uparrow\uparrow}+R_{\downarrow\uparrow} =1$ when $|E|<\Delta$.
In the absence of the Schottky potential, i.e., $Z=0$, the spin-flip reflection coefficient $R_{\downarrow\uparrow}$
is very large, where $R_{\downarrow\uparrow}$ is almost 1 for $|E|<\Delta$ and decays quickly for $|E|>\Delta$ (see Fig.\ref{fig8}(a)).
Since the Fermi wave lengths mismatch occurs at the two sides of the interface,
the spin-flip reflection is not perfect and $R_{\downarrow\uparrow}$ is slightly smaller than 1 at $Z=0$ and $|E|<\Delta$.
By increasing the barrier strength $Z$, the spin-flip reflection, the normal transmission, and the spin-flip
transmission are gradually reduced, whereas the normal reflection is strengthened. However, for $Z=0.3$ and $1.0$,
the spin-flip reflection remains considerably large [see Figs.\ref{fig8}(b) and \ref{fig8}(c)].
In addition, there always exists an incident energy $E$ between $0$ and $\Delta$ that $R_{\downarrow\uparrow} =1$ at this energy, regardless of the system's
parameters.

After obtaining the reflection and transmission coefficients, the particle current can be calculated
from the Landauer-B\"{u}ttiker formula\cite{sca3}
\begin{align}
I_{\sigma}=\frac{1}{h}[(R_{\bar{\sigma}\sigma} +
T_{\bar{\sigma}\sigma} +T_{\sigma\sigma})\mu_{L\sigma} -R_{\sigma\bar{\sigma}}\mu_{L\bar{\sigma}}].
\end{align}
Then, one can obtain the charge current $I_e$, the spin current $I_s$, as well as the differential charge conductance $G_e$ and
the differential spin conductance $G_s$ straightforwardly.

\begin{figure}[h]
\includegraphics[width=1.0\columnwidth]{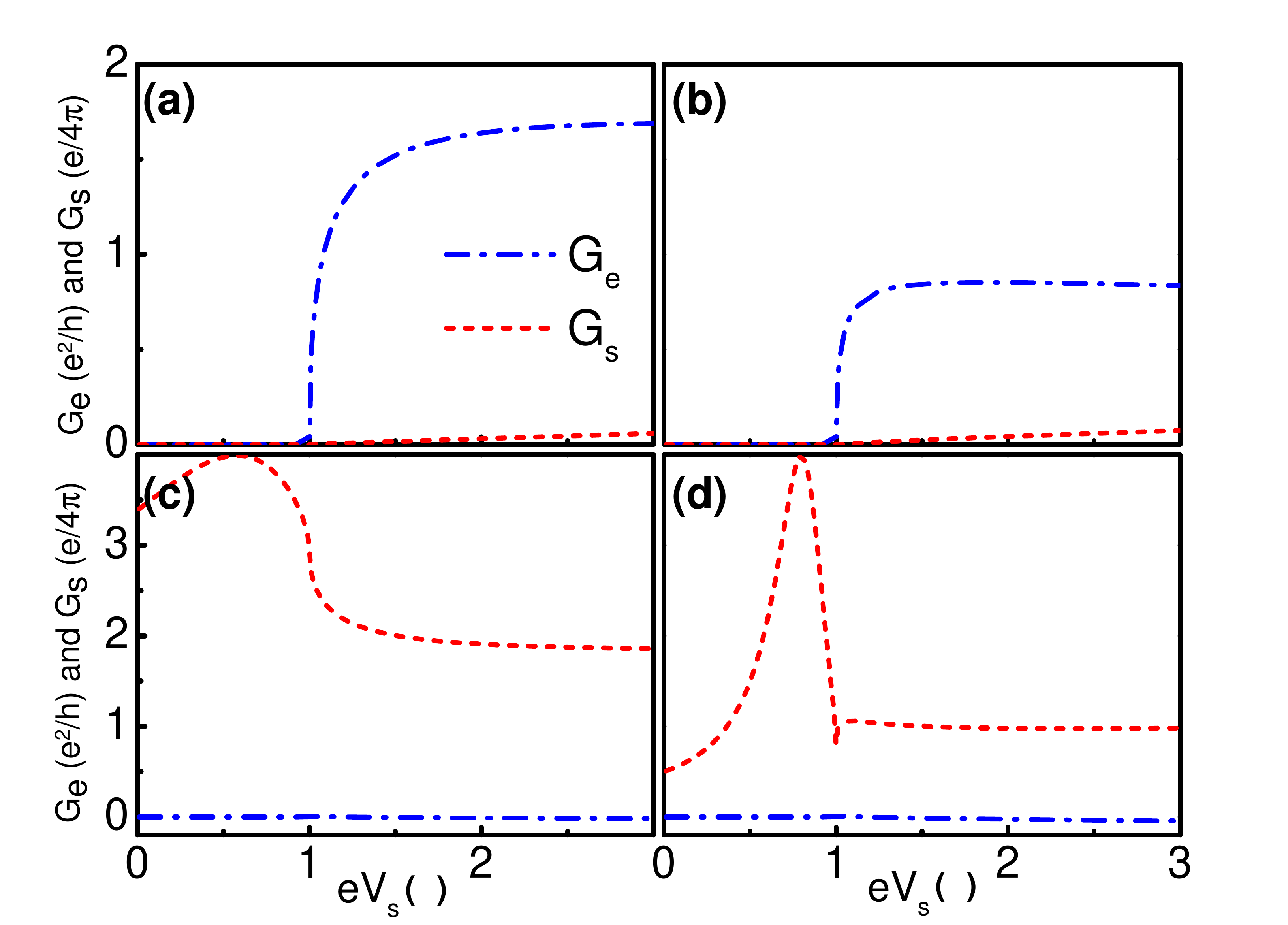}
\caption{(Color online) Differential charge conductance $G_{e}$ versus $eV_{s}$ and spin conductance $G_{s}$ versus $eV_{s}$
under a charge voltage [(a) and (b)] and under a spin voltage [(c) and (d)]. $Z=0.3$ in (a) and (c) and $Z=1.0$ in (b) and (d). The other parameters are the same as those in Fig.\ref{fig8}.} \label{fig9}
\end{figure}

Fig.\ref{fig9} shows the differential charge conductance $G_e$ and the
spin conductance $G_s$ under the charge voltage and the spin voltage.
In the case of the charge voltage, both $G_e$ and $G_s$ are zero when the voltage $eV_{s}$ is less than the gap $\Delta$
[see Figs.\ref{fig9}(a) and \ref{fig9}(b)],
which is similar to the graphene-based normal-spin superconductor junction.
Note that the spin superconductor is a charge insulator, and the charge voltage cannot drive the current.
When the voltage satisfies $eV_{s}>\Delta$, although the spin conductance $G_s$ is small, the charge conductance $G_e$ has a large value, due to the existence of the quasi-particle states above the gap.
In the case of the spin voltage,
the spin conductance $G_{s}$ has a large value even if the voltage is $eV_{s}<\Delta$ [see Figs.\ref{fig9}(c) and \ref{fig9}(d)]. For small barrier strengths $Z$, $G_s$ can exceed $2$ (in units of $e/4\pi$), because of the occurrence of the spin-flip reflection.
Although the potential $Z$ can suppress $G_s$, $G_s$ can always reach its maximum $4(e/4\pi)$ at a certain voltage
$eV_{s}$ below the gap $\Delta$, regardless of the potential $Z$.
These results indicate that the spin current can flow through the normal metal-spin superconductor junction due to
the spin-flip reflection.
In fact, the spin-flip reflection dominates the spin transport and
the spin conductance is proportional to the spin-flip reflection coefficient, with
$G_s =\frac{e}{2\pi}(R_{\downarrow\uparrow}+R_{\uparrow\downarrow})$
for $eV_{s}<\Delta$.
When $eV_{s}>\Delta$, the spin conductance $G_s$ can be also large because of the tunneling of the quasi-particle. Nevertheless, under the spin voltage,
the charge conductance $G_{e}$ is always very small.

\section{\label{sec:conclusions}Summary}

In summary, we study the spin-flip reflection at the normal metal-spin superconductor interface,
where a spin-up electron incident from the normal metal can be reflected as a spin-down one and a dissipationless super spin current can be generated in the spin superconductor.
We consider two cases in detail: the normal state has linear dispersion relation (graphene-based normal-spin superconductor junction) and quadratic dispersion relation.
For the graphene-based junction, a normal incident electron is perfectly spin-flip reflected at the interface,
in spite of the Fermi wave lengths mismatch at the two sides.
The oblique incidence slightly suppresses the spin-flip reflection and the Fermi wave lengths mismatch
begins to take effect.
In addition, the spin-flip reflection can be either specular reflection or
retro-reflection, depending on the incident and reflected electrons locating in the same band or the different bands.
As a comparison, the quadratic dispersion normal metal-spin superconductor junction is also studied,
where the spin-flip reflection coefficient can reach 1 at certain incident energies.
We calculate the differential charge conductance and the spin conductance under a charge voltage or a spin voltage.
In usual, the spin voltage can drive a large spin current through the junction, in which
the spin conductance is determined by the spin-flip reflection when the voltage is less than the gap of the spin superconductor.
These results will help us get a better understanding of the spin transport through the normal metal-spin superconductor junction.

\section*{Acknowledgement}

We gratefully acknowledge the support from NSF-China under Grant Nos. 11574007, 11274364, 11504066, and 11504008
NBRP of China under Grand No. 2015CB921102.


\section*{References}
\begin{thebibliography}{}
\bibitem{Andr}
A. F. Andreev, Sov. Phys. JETP \textbf{19}, 1228 (1964).

\bibitem{sca1}
G. E. Blonder, M. Tinkham, and T. M. Klapwijk, Phys. Rev. B \textbf{25}, 4515 (1982).

\bibitem{sca2}
C. W. J. Beenakker, Phys. Rev. B \textbf{46}, 12841 (1992).

\bibitem{sca3}
G. B. Lesovik and I. A. Sadovskyy, Physics-Uspekhi \textbf{54}, 1007 (2011).

\bibitem{spe1}
C. W. J. Beenakker, Phys. Rev. Lett. \textbf{97}, 067007 (2006).

\bibitem{spe2}
C. W. J. Beenakker, Rev. Mod. Phys. \textbf{80}, 1337 (2008).

\bibitem{addr1}
S.-G. Cheng, Y. Xing, J. Wang, and Q.-F. Sun, Phys. Rev. Lett. \textbf{103}, 167003 (2009).

\bibitem{gra1}
K. S. Novoselov, A. K. Geim, S. V. Morozov, D. Jiang, Y. Zhang, S. V. Dubonos,
I. V. Grigorieva, and A. A. Firsov, Science \textbf{306}, 666 (2004).

\bibitem{gra2}
K. S. Novoselov, A. K. Geim, S. V. Morozov, D. Jiang, M. I. Katsnelson, I. V. Grigorieva,
S. V. Dubonos, and A. A. Firsov, Nature \textbf{438}, 197 (2005).

\bibitem{gra3}
Y. Zhang, Y. W. Tan, H. L. Stormer, and P. Kim, Nature \textbf{438}, 201 (2005).

\bibitem{gra4}
A. H. Castro Neto, F. Guinea, N. M. R. Peres, K. S. Novoselov, and A. K. Geim,
Rev. Mod. Phys. \textbf{81}, 109 (2009).

\bibitem{gs1}
A. R. Akhmerov and C. W. J. Beenakker, Phys. Rev. B \textbf{75}, 045426 (2007);
D. Rainis, F. Taddei, F. Dolcini, M. Polini, and R. Fazio, Phys. Rev. B \textbf{79}, 115131 (2009).

\bibitem{gs2}
A. G. Moghaddam and M. Zareyan, Phys. Rev. B \textbf{74}, 241403(R) (2006).

\bibitem{gs3}
J. Linder and A. Sudb{\o}, Phys. Rev. Lett. \textbf{99}, 147001 (2007);
J. Linder and A. Sudb{\o}, Phys. Rev. B \textbf{77}, 064507 (2008).

\bibitem{gs4}
M. Maiti and K. Sengupta, Phys. Rev. B \textbf{76}, 054513 (2007).

\bibitem{gs5}
P. Burset, A. L. Yeyati, and A. Martin-Rodero, Phys. Rev. B \textbf{77}, 205425 (2008).

\bibitem{addr2}
S.-G. Cheng, H. Zhang, and Q.-F. Sun, Phys. Rev. B \textbf{83}, 235403 (2011).

\bibitem{addr3}
Y. Xing, J. Wang, and Q.-F. Sun, Phys. Rev. B \textbf{83}, 205418 (2011).

\bibitem{s1}
J. Bardeen, L. N. Cooper, and J. R. Schrieffer, Phys. Rev. \textbf{108}, 1175 (1957).

\bibitem{book1}
J. F. Annett, \textit{Superconductivity, Superfluids and Condensates} (Oxford University Press, Oxford, UK, 2004).

\bibitem{book2}
M. Tinkham, \textit{Introduction to Superconductivity} (Dover, New York, 2004).

\bibitem{book3}
A. J. Leggett, \textit{Quantum Liquids: Bose Condensation and Cooper Pairing in Condensed-Matter Systems}
(Oxford University Press, New York, 2006).

\bibitem{ssc1}
Q.-F. Sun, Z. T. Jiang, Y. Yu, and X. C. Xie, Phys. Rev. B \textbf{84}, 214501 (2011).

\bibitem{ssc2}
H. W. Liu, H. Jiang, X. C. Xie, and Q.-F. Sun, Phys. Rev. B \textbf{86}, 085441 (2012).

\bibitem{ssc3}
Q.-F. Sun and X. C. Xie, Phys. Rev. B \textbf{87}, 245427 (2013).

\bibitem{work1}
Z.-Q. Bao, X. C. Xie, and Q.-F. Sun, Nat. Commun. \textbf{4}, 2951 (2013).

\bibitem{lup1}
P. Lv, Z.-Q. Bao, A.-M. Guo, X. C. Xie, and Q.-F. Sun, Phys. Rev. B \textbf{95}, 014501 (2017).

\bibitem{He1}
Yu. M. Bunkov and G. E. Volovik, Phys. Rev. Lett. \textbf{98}, 265302 (2007).

\bibitem{He2}
M. Kupka and P. Skyba, Phys. Rev. B \textbf{85}, 184529 (2012).

\bibitem{NP1-111}
M.-S. Chang, Q. S. Qin, W. X. Zhang, L. You, and M. S. Chapman,
Nat. Phys. \textbf{1}, 111 (2005).

\bibitem{nature471-83}
Y.-J. Lin, K. Jim\'{e}nez-Garc\'{i}a, and I. B. Spielman, Nature \textbf{471}, 83 (2011).

\bibitem{PRL114-070401}
C. Hamner, Yongping Zhang, M. A. Khamehchi, Matthew J. Davis, and P. Engels,
Phys. Rev. Lett. \textbf{114}, 070401 (2015).

\bibitem{add1}
T. Giamarchi, C. R\"{u}egg, and O. Tchernyshyov, Nat. Phys. \textbf{4}, 198  (2008).

\bibitem{add2}
V. V. Mazurenko, M. V. Valentyuk, R. Stern, and A. A. Tsirlin, Phys. Rev. Lett. \textbf{112}, 107202 (2014).

\bibitem{add3}
S. Kimura, K. Kakihata, Y. Sawada, K. Watanabe, M. Matsumoto, M. Hagiwara, and H. Tanaka, Nat. Commun. \textbf{7}, 12822 (2016).

\bibitem{PRL68-1196}
U. Sivan, P. M. Solomon, and H. Shtrikman, Phys. Rev. Lett. \textbf{68}, 1196 (1992).

\bibitem{PRL73-304}
L. V. Butov, A. Zrenner, G. Abstreiter, G. B\"{o}hm, and G. Weimann, Phys. Rev. Lett.
\textbf{73}, 304 (1994).

\bibitem{spin-gra}
S. Takei, A. Yacoby, B. I. Halperin, and Y. Tserkovnyak, Phys. Rev. Lett. \textbf{116}, 216801 (2016).

\bibitem{Spin1}
S. Takei and Y. Tserkovnyak, Phys. Rev. Lett. \textbf{112}, 227201 (2014).

\bibitem{Spin2}
S. Takei, B. I. Halperin, A. Yacoby, and Y. Tserkovnyak, Phys. Rev. B \textbf{90}, 094408 (2014).

\bibitem{Spin3}
W. Chen and M. Sigrist, Phys. Rev. Lett. \textbf{114}, 157203 (2015).

\bibitem{twofluid}
B. Flebus, S. A. Bender, Y. Tserkovnyak, and R. A. Duine, Phys. Rev. Lett. \textbf{116}, 117201 (2016).

\bibitem{work2}
Q. Z. Zhu, Q.-F. Sun, and B. Wu, Phys. Rev. A \textbf{91}, 023633 (2015).

\bibitem{F1}
H. Haugen, D. Huertas-Hernando, and Arne Brataas, Phys. Rev. B \textbf{77}, 115406 (2008).

\bibitem{F2}
J. Linder, T. Yokoyama, D. Huertas-Hernando, and A. Sudb{\o}, Phys. Rev. Lett. \textbf{100}, 187004 (2008).

\bibitem{F3}
Q. Zhang, D. Fu, B. Wang, R. Zhang, and D. Y. Xing, Phys. Rev. Lett. \textbf{101}, 047005 (2008).

\bibitem{M1}
D. A. Abanin, S. V. Morozov, L. A. Ponomarenko, R. V. Gorbachev, A. S. Mayorov, M. I. Katsnelson, K. Watanabe, T. Taniguchi, K. S. Novoselov, L. S. Levitov, and A. K. Geim, Science \textbf{332}, 328 (2011).

\bibitem{H1}
M. Kharitonov, Phys. Rev. Lett. \textbf{109}, 046803 (2012);
M. Kharitonov, Phys. Rev. B \textbf{86}, 075450 (2012).

\bibitem{H2}
Y. Hama, G. Tsitsishvili, and Zyun F. Ezawa, Phys. Rev. B \textbf{87}, 104516 (2013).

\bibitem{addr4}
D.-K. Wang, Q.-F. Sun, and H.Guo, Phys. Rev. B \textbf{69}, 205312 (2004).

\bibitem{addr5}
Q.-F. Sun, H. Guo, and J. Wang, Phys. Rev. Lett. \textbf{90}, 258301 (2003);
L. Wen, Q.-F. Sun, H. Guo, and J. Wang, Appl. Phys. Lett. \textbf{83}, 1397 (2003).

\bibitem{sv11}
S. O. Valenzuela and M. Tinkham, Nature \textbf{442}, 176 (2006).

\bibitem{sv12}
N. Tombros, C. Jozsa, M. Popinciuc, H. T. Jonkman, and B. J. van Wees, Nature \textbf{448}, 571 (2007).

\bibitem{sv13}
S. M. Frolov, S. L$\ddot{u}$scher, W. Yu, Y. Ren, J. A. Folk, and W. Wegscheider, Nature \textbf{458}, 868 (2009);
S. M. Frolov, A. Venkatesan, W. Yu, J. A. Folk, and W. Wegscheider, Phys. Rev. Lett. \textbf{102}, 116802 (2009).

\bibitem{sv1}
K. Uchida, S. Takahashi, K. Harii, J. Ieda, W. Koshibae, K. Ando, S. Maekawa, and E. Saitoh, Nature \textbf{455}, 778 (2008);
K. Uchida, J. Xiao, H. Adachi, J. Ohe, S. Takahashi, J. Ieda, T. Ota, Y. Kajiwara, H. Umezawa, H. Kawai G. E. W. Bauer, S. Maekawa, and E. Saitoh, Nature Mater. \textbf{9}, 894 (2010).

\bibitem{sv2}
C. M. Jaworski, J. Yang, S. Mack, D. D. Awschalom, J. P. Heremans, and R. C. Myers, Nature Mater. \textbf{9}, 898 (2010);
C. M. Jaworski, R. C. Myers, E. Johnston-Halperin, and J. P. Heremans, Nature  \textbf{487}, 210 (2012).

\bibitem{sv3}
S. M. Wu, W. Zhang, K. C. Amit , P. Borisov, J. E. Pearson, J. S. Jiang, D. Lederman, A. Hoffmann, and A. Bhattacharya,
Phys. Rev. Lett. \textbf{116}, 097204 (2016).

\bibitem{newnew}
F. Bottegoni, M. Celebrano, M. Bollani, P. Biagioni, G. Isella, F. Ciccacci and M. Finazzi, Nat. Mater. \textbf{13}, 790 (2014).
\end{thebibliography}
\end{document}